\documentclass[prb,twocolumn,showpacs,preprintnumbers,amsmath,amssymb]{revtex4}
\usepackage{amssymb}
\usepackage{bm}
\usepackage{amsmath}
\usepackage[dvips]{graphicx}
\usepackage{color}

\begin{document}

\title{A one-dimensional spin-orbit interferometer}
\author{T. Li}
\affiliation{School of Physics, University of New South Wales, Sydney 2052, Australia}
\author{O. P. Sushkov}
\affiliation{School of Physics, University of New South Wales, Sydney 2052, Australia}

\begin{abstract}
We demonstrate that the combination of an external magnetic field and the intrinsic spin-orbit interaction
 results in nonadiabatic precession of the electron spin after transmission
through a quantum point contact (QPC).
We suggest that this precession may be observed in a device consisting of two QPCs placed in series. 
The pattern of resonant peaks in the transmission is strongly influenced by the non-abelian 
phase resulting from this precession.
Moreover, a novel type of resonance which is associated with suppressed, rather than enhanced,
transmission emerges in the strongly 
nonadiabatic regime.
The shift in the resonant transmission peaks is dependent on the spin-orbit interaction and therefore
offers a novel way to directly measure these interactions in a ballistic 1D system.
\end{abstract}

\maketitle

\section{Introduction}

In recent years, it has been recognized that the existence of interactions which couple spin to orbital 
motion gives rise to the possibility of manipulating the spin via external gates, leading to the 
suggestion of spintronic devices which require only electric fields for their operation \cite{DattaDas}. 
The importance of the \emph{spin-orbit interactions} lies both in their role in spin dephasing via the 
Dyakonov-Perel \cite{SR:DyakonovPerel1,SR:DyakonovPerel2} and Elliot-Yafet \cite{SR:Elliott,SR:Yafet} 
mechanisms, as well as in the potential creation of spin-polarized 
current \cite{SF:Eto,SF:Gelabert,SF:GovernaleZulicke2003,SF:GovernaleZulicke2004,SF:Kita2004,SF:LiuZhou,SF:Xiao}.

In the present work we investigate the possibility of the dynamical manipulation of spin via a 
combination of electric and magnetic fields in a one-dimensional system. We consider the interactions 
which couple to the first power of spin and therefore have the structure of a magnetic dipole. In the 
literature, the dominant interaction of this form is usually considered to be due to the inversion 
asymmetry of the two-dimensional (2D) interface, known as the \emph{Rashba effect}, while 
the \emph{Dresselhaus effect} arising from inversion asymmetry in the bulk crystal has been considered 
to be negligibly small. While for the narrow-gap semiconductors InAs and InSb, the Rashba interaction is 
dominant, the coefficient of the Rashba interaction varies by two orders of magnitude between the 
narrow-gap and medium-gap materials \cite{WinklerBook} and we expect that in GaAs the situation is reversed. 
It has previously been possible to determine the relative size of the Rashba and Dresselhaus interactions 
in 2D GaAs via Faraday rotation \cite{MeierSalis}, where they were determined to be approximately equal in 
magnitude. 
The Dresselhaus interaction was also found to be approximately  constant over a range of samples, which 
is puzzling since it is expected to scale quadratically with the width of the quantum well, as was 
noted by the authors of \cite{MeierSalis}.

The dipolar structure of the Rashba and Dresselhaus interactions determines effective magnetic fields 
which, in the presence of an external magnetic field, will align spin along the direction given by 
their vector sum with the external field. As existing measurements of the spin-orbit interaction in 
2D systems rely on diffusive transport when both the Rashba and Dresselhaus interactions are proportional 
to the very small average momentum resulting in a small effective
magnetic field of order of 1mT. 
In contrast, in a ballistic system, the spin-orbit interactions are proportional to the Fermi momentum 
$k_F$, which is several orders of magnitude higher than the average momentum in the diffusive regime.
We therefore suggest a novel method to measure the spin-orbit interactions in a quantum wire which relies 
on spin-orbit induced nonadiabaticity inside a ballistic channel.

For simplicity, we  consider a quantum point contact formed from a 2D {\it electron} gas with only 
the Dresselhaus interaction present, although we also found numerically that hole systems show similar 
behaviour. For a wire oriented along the $x  = (100)$ direction, we find upon projection of the bulk 
Hamiltonian onto one-dimensional (1D) states,
\begin{gather}
\label{hd}
H_D = b^{6c6c}_{41} (p_x(p_y^2 - p_z^2)\sigma_x + p_y(p_z^2 - p_x^2) \sigma_y + \\ \nonumber
 p_z(p_x^2 - p_y^2) \sigma_z) 
\rightarrow b^{6c6c}_{41}p_x(p_y^2 - p_z^2)\sigma_x \ ,
\end{gather}
where $b^{6c6c}_{41} \approx 28 \text{eV}\AA^{3} \hbar^{-3}$ is the Dresselhaus constant \cite{WinklerBook}, 
and we have set $\langle p_y \rangle = \langle p_z \rangle = 0$, assuming that 
the $y$- and $z$ confinements are along (010) and (001) respectively. Here $\sigma_i$ are the Pauli matrices
describing spin.

\begin{figure}[ht]
\includegraphics[width=0.23\textwidth,clip]{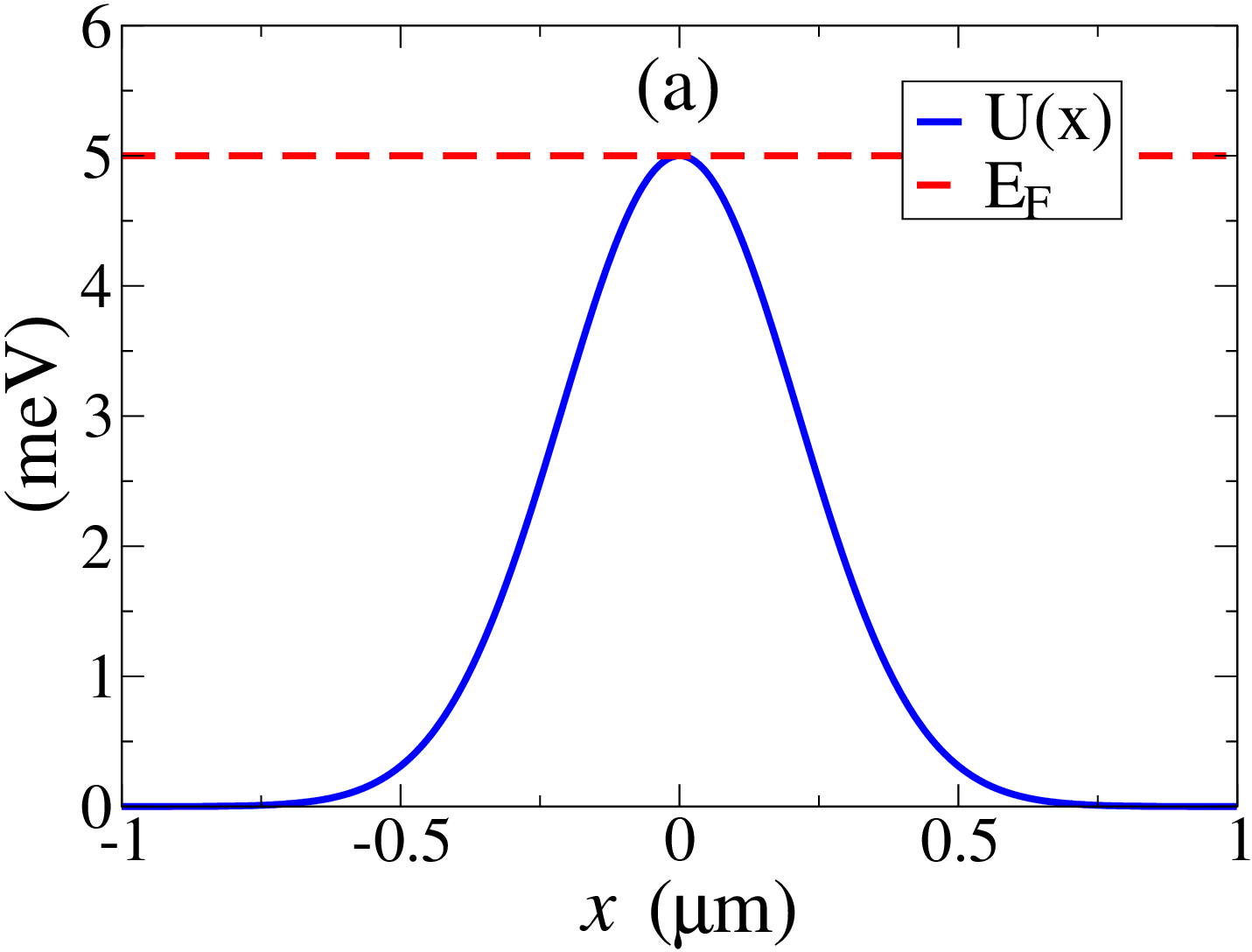}
\includegraphics[width=0.23\textwidth,clip]{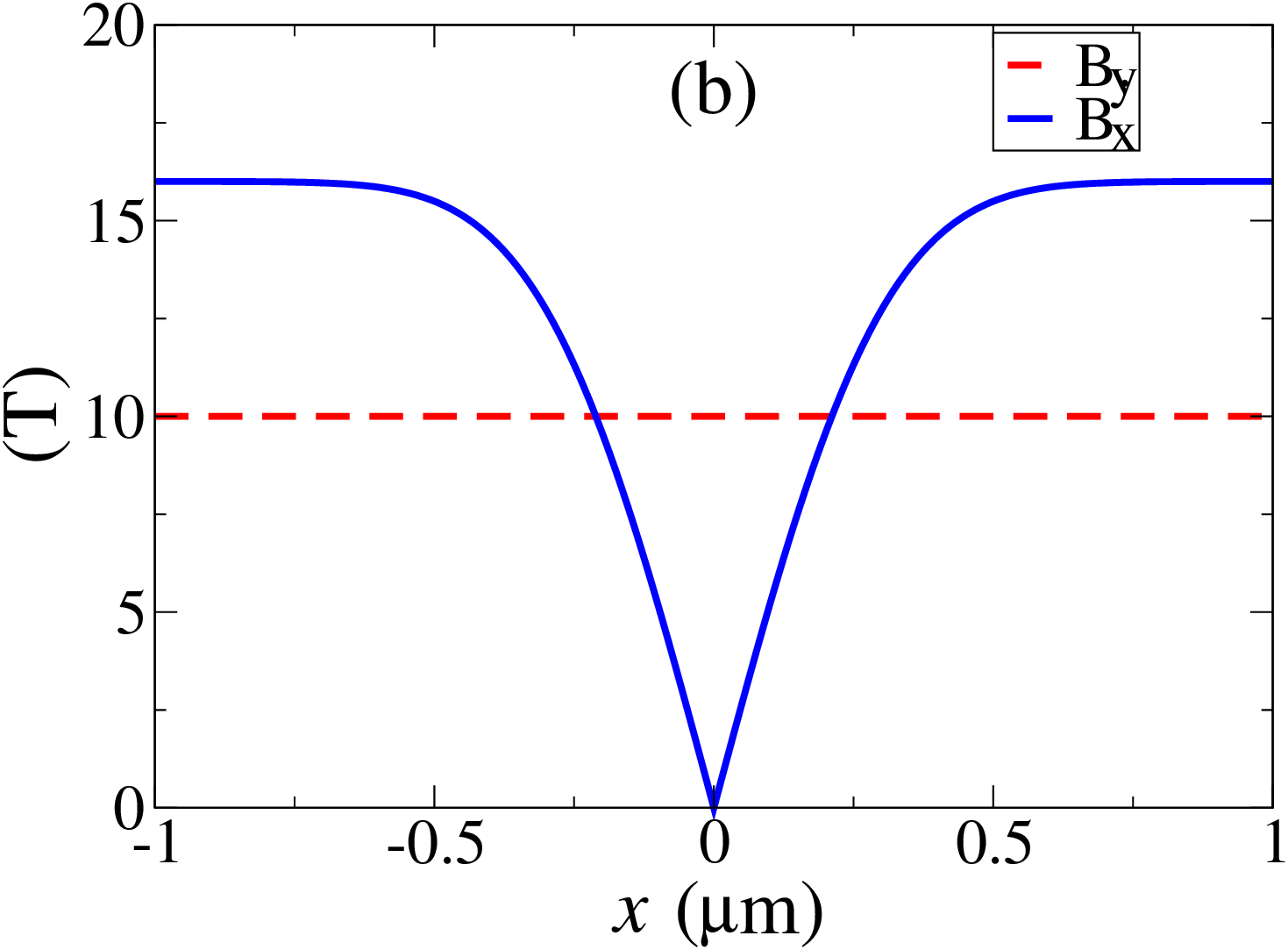}
\caption{\emph{(Color online).
{\bf (a)} The solid blue line shows the effective potential in a single QPC, modelled as a gaussian,
the length of the channel is 2$\mu$m. The dashed red line shows the Fermi energy. 
{\bf (b)} The components of the effective magnetic field parallel (blue, solid) and perpendicular 
(red, dashed) to the QPC.
The perpendicular component is  the external magnetic field. 
The parallel component is the Dresselhaus effective field, which decreases semiclassically at the top 
of the potential shown in the panel {\bf (a)}.
}}
\label{fig:scatteringbarrier}
\end{figure}

According to Eq. (\ref{hd}) the Dresselhaus magnetic field is parallel to the wire. The field is inhomogeneous 
in the presence of an electrostatic barrier, Fig.\ref{fig:scatteringbarrier}a,
since the effective magnetic field is proportional to the momentum which vanishes semiclassically at the 
turning points. When an external magnetic field is applied perpendicular to the contact, the combination 
of the Zeeman interaction and the Dresselhaus interaction forms a total driving torque on spin which is 
inhomogeneous in space and rapidly switches direction, see Fig.\ref{fig:scatteringbarrier}b. 
We propose that the non-trivial spin dynamics resulting from the inhomogeneous driving field may be observed 
in a double barrier interference experiment, and such an experiment will distinguish between adiabatic and 
nonadiabatic spin motion and therefore serve as a direct measurement of the size of the Dresselhaus 
interaction. The same logic is applicable to the Rashba interaction.

The structure of the paper is as follows:
In Section II we formulate the concept of nonadiabatic spin precession and present the criteria for its existence,
based on typical experimental parameters.
In Section III we introduce the idea of an interferometer consisting of a double quantum point contact (DQPC),
and the discuss the adiabaticity of the ``orbital" dynamics which is required for
interference to be observed.
In Section IV we present results of numerical solution of the Schrodinger equation describing
electron transmission through the interferometer in the presence of external magnetic and spin-orbit fields, and demonstrate
how measurement of the Dresselhaus/Rashba interaction can be performed in the device.
In Section V we present the physical mechanism behind the phenomenon of ``negative" resonances which are observed in the numerical result
and show that they are a strong signature of nonadiabatic spin dynamics.
In Section VI we present our conclusions.

\section{nonadiabaticity due to spin}
We consider a one-dimensional channel formed by electrostatic confinement in a 2D electron gas (2DEG) in GaAs,
where the 2D quantum well is grown along the (100) direction. In the presence of an external magnetic field 
and the Dresselhaus interaction the conductance is determined by the solution of the spin-dependent 
transmission problem. The effective Hamiltonian for a single channel reads
\begin{gather}
H = \frac{p_x^2}{2m} + U(x) -\frac{1}{2} g^* \mu_B \vec{B}\cdot \vec{\sigma} +b^{6c6c}_{41} p_x ( p_y^2  - p_z^2) \sigma_x \ ,
\end{gather}
where $U(x)$ is the electrostatic barrier,
for GaAs, $m = 0.067 m_e$ is the effective mass and $g^* = -0.44$ the effective $g$-factor \cite{WinklerBook}. $p_y^2$ and  $p_z^2$ refer to the average of the differential operators in the bound states 
along $y$ and $z$ respectively. We shall assume the 2D limit, so that $p_y^2 \ll p_z^2$.

\begin{figure}[ht]
\includegraphics[width=0.23\textwidth,clip]{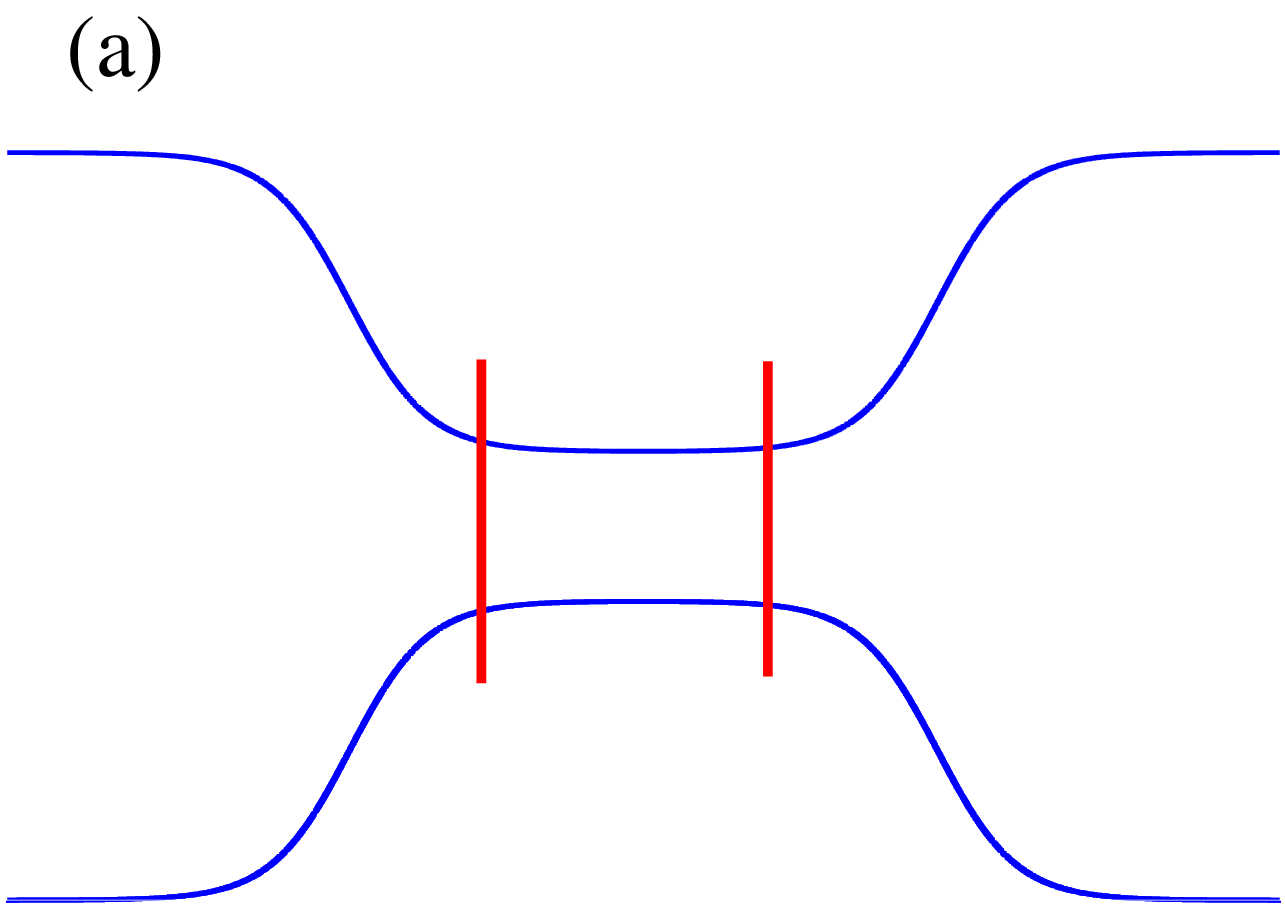}
\includegraphics[width=0.23\textwidth,clip]{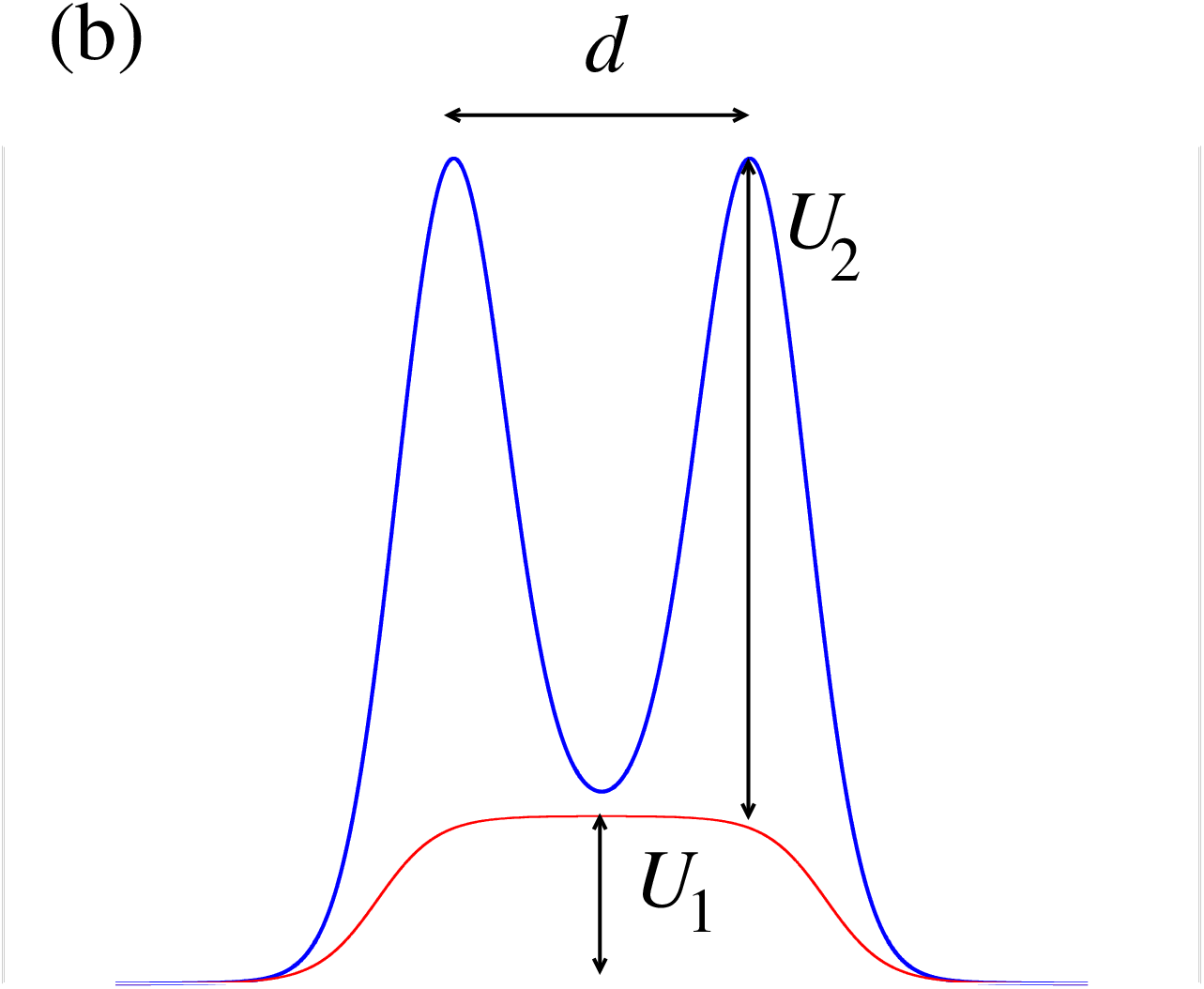}
\caption{\emph{(Color online).
(a) A DQPC formed from a ballistic quantum wire. The double barrier potential is created by two narrow wires above the 1D channel. (b) The effective 1D potential, which consists of the potential due to gates $U_1$ and additional gaussian barriers of height $U_2 = 4\text{meV}$ separated by a distance $d$, which is fixed at 2$\mu$m. The contribution $U_1$ to the effective potential (red) due to side gates is approximately constant between the barriers.}}
\label{fig:dqpc}
\end{figure}
Let us consider the polarization of the asymptotic states. We will assume that only a few transverse channels 
are open in the QPC, and in the highest channel, the Fermi energy is close to the top of the barrier. 
In the asymptotic region, the confinement along $y$ becomes infinitely wide as the channel smoothly 
connects to the 2D leads, Fig.\ref{fig:dqpc}a, so that for scattering states at the Fermi energy,
\begin{gather}
p_y^2, U(x) \rightarrow 0 \\ \nonumber
p_x \rightarrow p_F
\end{gather}
and hence
\begin{gather}
H(x \rightarrow \pm \infty) = \frac{ p_x^2}{2m} - \frac{1}{2} | g\mu_B| \vec{B}_{\text{eff.}}\cdot \vec{\sigma}\ ,
\end{gather}
where we have absorbed the total spin-dependent interactions into an effective magnetic field
\begin{gather}
\vec{B}_{\text{eff.}}(\pm \infty) = \vec{B} + \frac{ b^{6c6c}_{41} p_F p_z^2}{\frac{1}{2} | g\mu_B |}  \hat{x} \ .
\end{gather}
For the purpose of numerical calculations which we present in Section IV, we assume that the value of $p_z$ 
corresponds to an infinite well with width 10nm and set the Fermi energy equal to 5meV. Based on the values 
given in \cite{WinklerBook}, we may estimate the effective Dresselhaus field to be
\begin{gather}
B_D = \frac{ b^{6c6c}_{41} p_F p_z^2}{\frac{1}{2} | g^*\mu_B |} = 16\text{T}\ .
\end{gather}
When the external magnetic field is directed along $y$, the asymptotic form of the Hamiltonian will form a 
spin basis with polarization the $x-y$ plane, and for fields in the typical experimental range, 
$B < \text{15T}$, the orientation of spin for an electron incident on the barrier will always be 
significantly tilted towards the $x$-direction. The  angle $\theta$ between spin and the $y$-direction is 
close to 60$^0$ for fields of the order of 10T, and can be tuned to nearly 90$^0$ by reducing the field to 
1T. Near the centre of the QPC, where the electrostatic potential is maximum, the 
longitudinal momentum vanishes semiclassically, and the total effective magnetic field is directed 
perpendicular to the wire, so that $\theta$ switches to zero, Fig.\ref{fig:scatteringbarrier}b.

Spin dynamics is nonadiabatic when the effective magnetic field changes sufficiently rapidly and hence
the Landau-Zener parameter is not small~\cite{com1},
\begin{gather} 
\label{eq:LZ}
\delta_s = \frac{1}{g^* \mu_B B_{\text{eff.}}^2} |\frac{ dB_{\text{eff.}} }{ dt} | \gtrsim 1
\end{gather}
Let us assume that the effective field $B_{\text{eff.}}$ switches by an angle of $60^0$ over a typical time 
$\Delta t = \frac{\Delta x}{v_F}$ corresponding to the distance $\Delta x$ over which the electrostatic potential 
is rapidly varying. Then expressing $\delta_s$ in terms of the Fermi energy $\epsilon_F \approx 5\text{meV}$, we find
\begin{gather}
\delta_s =\frac{1}{g \mu_B |B_{\text{eff.}}|} \frac{v_F}{\Delta x}  \cos 60^0 
 = \frac{1}{2} \frac{\epsilon_F}{ g\mu_B |B_{\text{eff.}|}}  \frac{1}{\Delta x k_F}\ .
\end{gather}
Taking $|B_{\text{eff.}}| \approx 10\text{T}$, we find that $ \frac{ \epsilon_F}{ g\mu_B |B_{\text{eff.}}|} \approx 20$, so in order to go to the nonadiabatic regime one needs the following,
\begin{gather}
\Delta x k_F \ll 10 \rightarrow \Delta x \lesssim 0.1\mu\text{m}\ .
\end{gather}

For a single barrier, the conductance is proportional to the transmission coefficient averaged over 
incident spin polarizations and is therefore insensitive to spin dynamics~\cite{com2}. 
In a geometry consisting of two barriers, however, non-trivial precession between the barriers may be 
observed due to interference between spin in counterpropagating directions in the region between the 
barriers. This effect exists only when the external field is perpendicular to the contact, since for 
a parallel field, the total effective field and therefore spin will be constant in direction, 
either parallel or antiparallel to the contact with no mixing between the spin modes. 
Similarly, when the external field is perpendicular but sufficiently large to dominate the effective 
field  in the asymptotic region, the effective field will rotate by a sufficiently small angle that 
the electron  will adiabatically follow a single spin channel throughout the motion. This illustrates 
why the structure  of transmission in the double barrier is strongly sensitive to the 
nonadiabaticity of the spin motion.

\section{Adiabatic transmission through a double barrier}

Let us first consider the spin-independent transmission problem for a system consisting of two 
QPCs in series, with the potential as shown in Fig.\ref{fig:dqpc}b.
We shall assume that the inelastic mean free path exceeds the system size, so that motion is ballistic.

We model the conductance in the Landauer-B\"{u}ttiker picture \cite{ButtikerQPC,Landauer} by 
one-dimensional scattering in the presence of two barriers separated by a distance $d$. 
Recall that for a \emph{single barrier} located at the origin we have a scattering state 
corresponding to an electron emerging on the right with unit amplitude,
\begin{gather}
\psi(x \gg 0) = e^{i k x} \\ \nonumber
\psi(x \ll 0) = t_{11} e^{i kx} + t_{21} e^{-ikx} \ .
\end{gather}
Near the top the barrier has parabolic shape
\begin{gather}
U(x \approx 0) = U(0) - \frac{ m \omega_x^2 x^2}{2}
\end{gather}
and hence  $T$-matrix elements $t_{11}, t_{21}$ are given by the connection formulas of the 
parabolic cylinder functions \cite{AbramowitzStegun},
\begin{gather} 
\label{tmatrix}
T = \left(\begin{array}{cc}
t_{11} & t_{21}^*\\
t_{21} & t_{11}^*
\end{array}\right)  = \\ \nonumber
 \left(\begin{array}{cc}
i \sqrt{ 1 + e^{ -\frac{ 2\pi \epsilon}{\omega_x} } } & i e^{- \frac{\pi \epsilon}{\omega_x} -i \text{Arg} (\frac{1}{2} - \frac{i\epsilon}{\omega_x}) } \\
-i e^{-\frac{\pi \epsilon}{\omega} + i\text{Arg} (\frac{1}{2} -\frac{i\epsilon}{\omega_x} ) } & -i \sqrt{ 1 + e^{- \frac{ 2\pi \epsilon}{\omega_x}} }
\end{array}\right) \ ,
\end{gather}
where $\epsilon$ is the Fermi energy relative to the height of the barrier,
\begin{gather}
\epsilon = \varepsilon_F - U(0) \ .
\end{gather}
Note that for the spinless case the T-matrix has dimension 2$\times$2.
The transmission probability through the single barier is
\begin{gather}
P(\epsilon)=\frac{1}{|t_{11}|^2}
\end{gather}

The $T$-matrix for transmission through the double barrier potential is given by the product of two 
$T$-matrices for a single barrier with the phase evolution operator,
\begin{gather}
T' = T \left(\begin{array}{cc} e^{i k d} & 0 \\ 0& e^{-i k d} \end{array}\right)T \ .\\ \nonumber
\end{gather}
Since the state emerging on the right has unit amplitude, the transmission probability is simply given by
\begin{gather}
P(\epsilon) = \frac{1}{|t'_{11}|^2} = \frac{1}{|t_{11}^2 + |t_{21}|^2 e^{2i kd }|^2} =  \\ \nonumber
\frac{1}{|1 + e^{- \frac{ 2\pi \epsilon}{\omega_x}} (1 - e^{2i kd})^2 |^2}
\end{gather}
and exhibits resonant transmission peaks at $ kd = n \pi $
corresponding to standing waves between the  barriers.

Initial measurements of the quantised conductance in a DQPC by D. Wharam, 
\emph{et al} \cite{Wharam} did not reveal resonant structure, although it was later reported by 
Y. Hirayama and T. Saku that resonances became visible when the separation between the two 
QPCs was reduced to 0.2$\mu$m \cite{Hirayama}. Since the authors claim that the inelastic mean free 
path exceeds the size of the device, the loss of interference does not originate from inelastic 
decoherence, but is rather due to the fact that the region between the QPCs consists of a wide 
cavity in which a large number of transverse modes are permitted. Here, the loss of phase memory 
may be attributed to the exchange of phase among the large number of transverse modes and is 
therefore a purely single particle effect. 
In other words, phase memory is lost in the course of chaotic motion in the 2D region separating 
the QPCs.

In order to observe the transmission interference peaks, it is necessary to suppress mixing between 
transverse modes in the region between the barriers, which is equivalent to the statement that the 
evolution of the standing wave along $y$ must be \emph{adiabatic}, and therefore the Landau-Zener 
condition must be satisfied for the \emph{transverse adiabatic parameter.}
\begin{gather}
\delta_t = \frac{1}{\omega_y^2} \frac{ d \omega_y}{dt} = \frac{1}{\omega_y^2} \frac{p_x}{m} \frac{ d\omega_y}{dx} \ll 1 \ ,
\end{gather}
where we have assumed a parabolic confinement in the $y$-direction with level spacing $ \omega_y$. 
For a typical QPC, the oscillator frequency $\omega_y$ is maximum at the centre of the wire ($x = 0$), 
but decreases smoothly to zero in the two-dimensional leads. Modelling the transverse confinement
by a gaussian,
\begin{gather}
\omega_y(x) = \omega_0 e^{- \kappa^2 x^2} \ ,
\end{gather}
where $\kappa $ is the barrier width, which is of order $\kappa \sim 1 \mu m$, we find
\begin{gather} \label{divergentgamma}
\delta_t = -\frac{2 \kappa^2 x p_x}{m \omega_0} e^{\kappa^2 x^2}\ . 
\end{gather}
Away from the contact, $x \rightarrow \pm \infty$, the adiabatic parameter diverges due to the collapsing 
of the transverse level spacing.
We therefore see that a loss of interference is unavoidable in a system consisting of two QPCs which are separated
by a wide cavity. Hereafter we consider an interferometer which consists of 
a 1D channel of fixed width in which a double barrier is formed by an additional potential $U_2(x)$ (\emph{i.e.} 
an inhomogeneous shift of the 1D band bottom) rather than by the energy of transverse confinement. 
In the region between the barriers, the 2D potential has the form
\begin{gather} \label{potentialhumps}
U_{2D}(x, y) = \frac{ m \omega_y^2 y^2 }{2} + U_2(x) \ ,
\end{gather}
and the oscillator frequency $\omega_y$ is approximately constant \emph{inside} the channel, so that an electron 
remains in a single transverse mode during motion between the barriers.

The potential (\ref{potentialhumps}) can be manufactured, for example, by a rectangular split gate in 
the plane of the 2DEG, with two thin wires placed perpendicular to the channel in a plane separated 
from the 2DEG by an insulating layer, Fig. \ref{fig:dqpc}a. When there is a bias between the wires in 
the upper layer and the 2DEG, a smooth electrostatic potential will be formed in the channel below. The 
1D channel is quantized into oscillator levels, with the 1D effective potential being
\begin{gather} \label{U2}
U(x) = U_1(x) + U_2(x) \\ \nonumber
U_1(x) = \omega_y(x) (n + \frac{1}{2})
\end{gather}
where $n$ is the transverse oscillator level. We also suppose that the channel is not near pinch-off, 
so that the additional barrier $U_2(x)$ may be made high without depleting the channel. As long as the 
wires are placed inside the edges of the point contacts, the level spacing will be constant, and 
transport between the potential barriers in the channel created by the wires will be adiabatic. 

\section{Resonant transmission of spin}

We expect that the presence of nonadiabatic spin precession will result in an observable change in the 
conductance when the distance between the barriers is of the order of the length of a spin cycle,
\begin{gather}
l_s = \frac{2\pi v_F}{\frac{1}{2} g^*\mu_B B} = \frac{ \epsilon_F}{\frac{1}{2} g^*\mu_B B} \frac{2\pi }{k_F} = \frac{\epsilon_F}{\frac{1}{2} g^*\mu_B B} \lambda = 1.2\mu\text{m}
\end{gather}
when the Fermi energy is 5meV. We reiterate that to observe the effect of nonadiabaticity on the conductance, 
it is necessary to have $\delta_s \gg 1$ at the barriers, and $\delta_t \ll 1$ everywhere in the region between 
the barriers.

Due to the small $g$-factor of electrons, the longitudinal oscillator frequency $\omega_x$ must be  
sufficiently small in order to resolve the Zeeman splitting in the \emph{external magnetic field}, and we 
take the value of $\omega_x = 0.19\text{meV}$ corresponding to a gaussian half-width of $0.4\mu$m. 
When $d < 2\mu$m, we find that the potentials of the two barriers overlap, reducing the velocity and hence 
the magnitude of the Dresselhaus effective field between the barriers. In principle, it is possible to engineer 
a system with $d = 1\mu$m while maintaining a large Dresselhaus field by using a third wire above the point 
contact which is positively biased to create a deeper cavity between the barriers. In our numerics, however, 
we will consider only the simpler geometry consisting of two wires in a plane above the point contact, and 
take a larger separation, $d = 2\mu$m, see Fig.\ref{fig:dqpc}b.
The 1D Hamiltonian is
\begin{gather} 
\label{simpHamil}
H = \frac{p_x^2}{2m} + U(x) -\alpha p_x \sigma_x - \frac{1}{2} g^* \mu_B \vec{B} \cdot \vec{\sigma} \ ,
\end{gather}
where $\alpha = b^{6c6c}_{41} p_z^2$ and we have dropped the term proportional to $p_y^2$, which
 does not qualitatively influence the result.
When the external magnetic field is directed along $x$, the Hamiltonian 
becomes diagonal in the basis of states with spin aligned along the contact, and the transmission coefficients 
for the spin-up and spin-down channels are simply given by the sum of two zero-field transmission coefficients 
shifted relative to one another by the Zeeman splitting $|g^* \mu_B B|$; the Dresselhaus interaction does not 
have any effect. For the perpendicular orientation of the external field, the behaviour of the transmission 
coefficient is expected to be significantly more complex due to nonadiabaticity, since the Hamiltonian does 
not decouple in any locally defined basis.

\begin{figure}[ht]
\includegraphics[width=0.23\textwidth,clip]{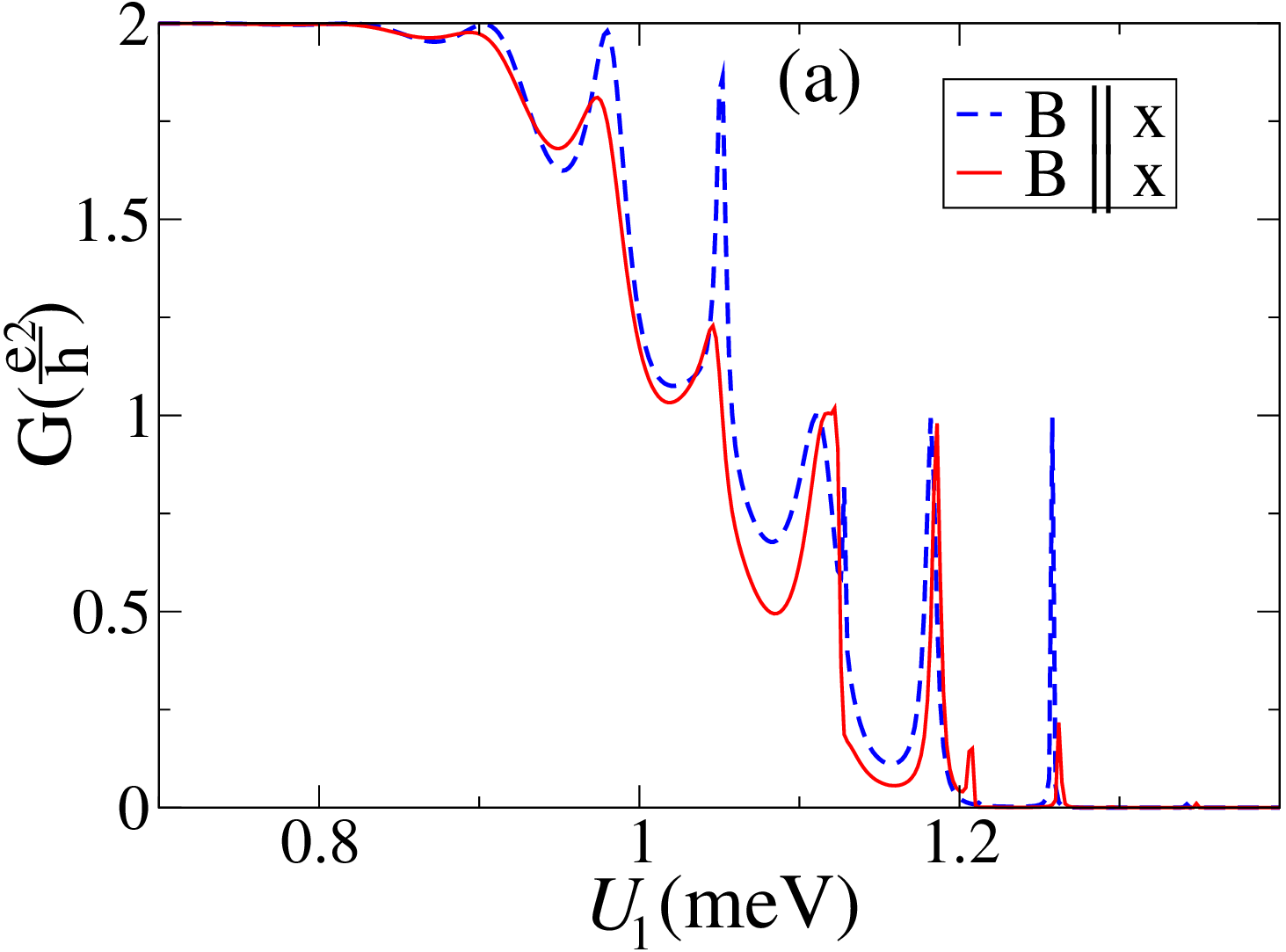}
\includegraphics[width=0.23\textwidth,clip]{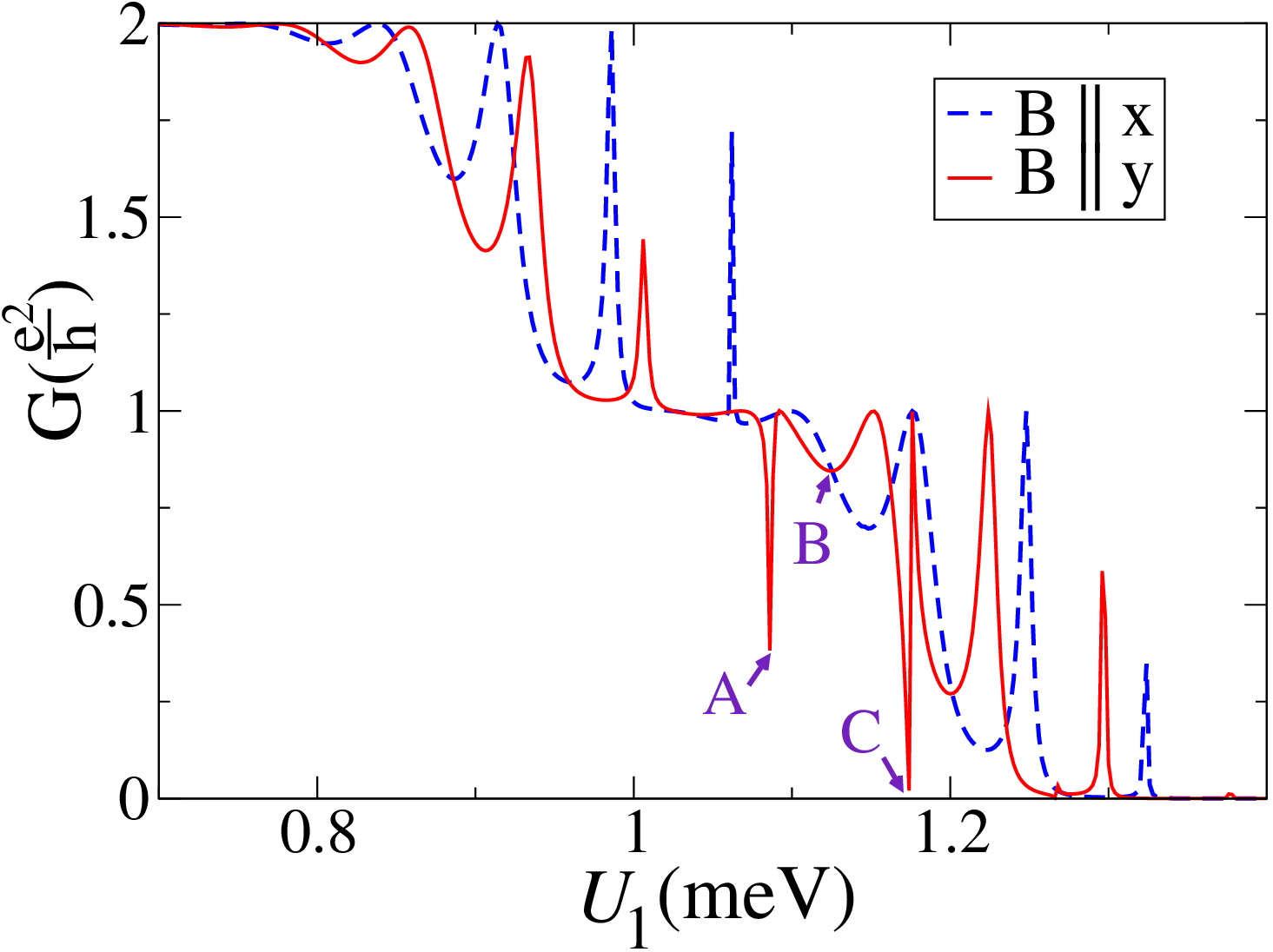}
\includegraphics[width=0.23\textwidth,clip]{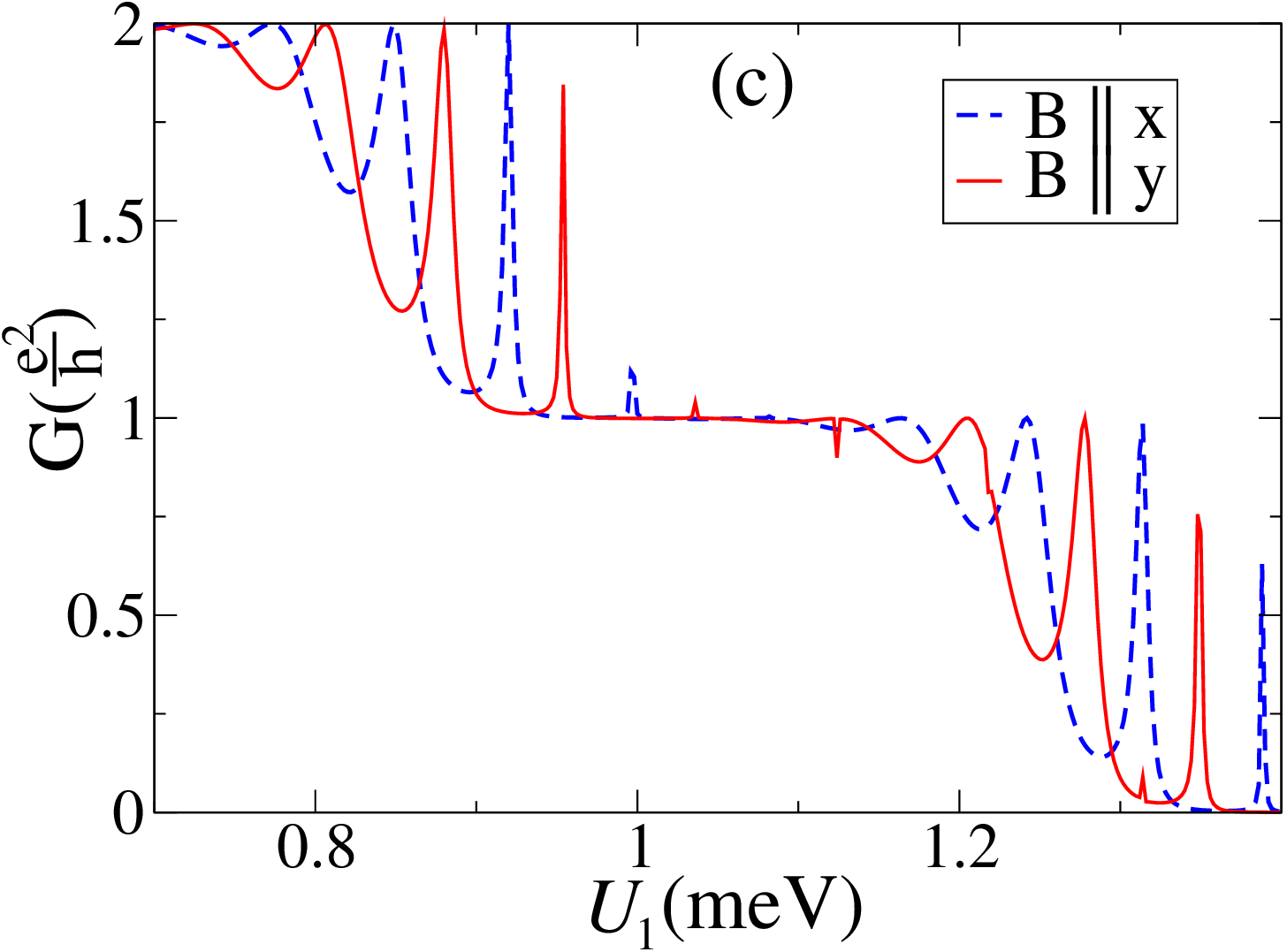}
\caption{\emph{(Color online).
The conductance (units of $\frac{e^2}{h}$) of the DQPC interferometer ($d=2\mu$m)
versus the constant contribution to the potential $U_1$ defined in (\ref{U2}).  The conductance is plotted for three values of the external magentic field,
(a)5T, (b) 10T, (c) 15T, parallel (dotted line) and perpendicular (solid line) to the contact. 
The arrows in (b) refer to the three plots in Fig.\ref{fig:boundstate}.}}
\label{fig:conductance}
\end{figure}

\begin{figure}[ht]
\includegraphics[width=0.23\textwidth,clip]{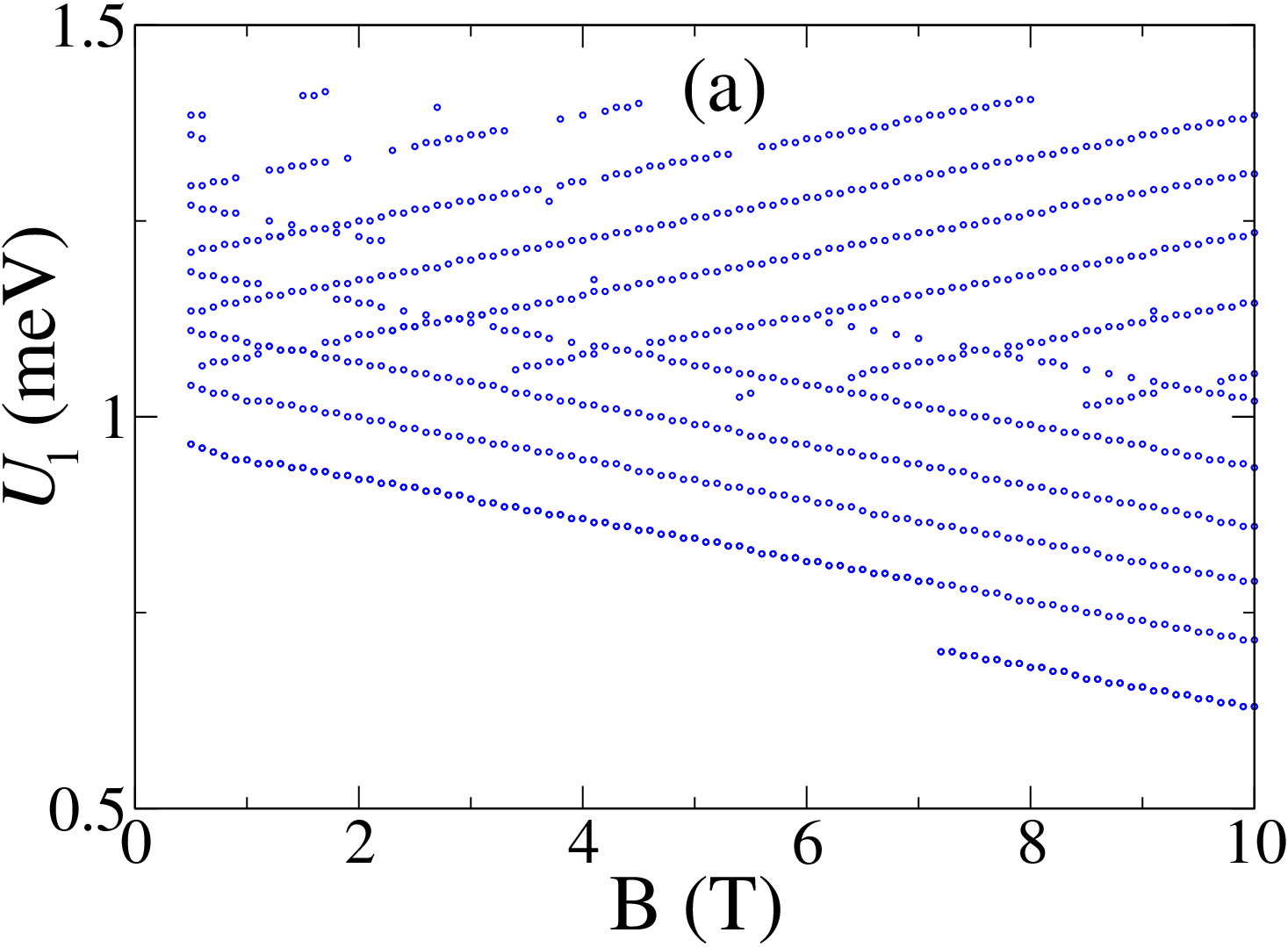}
\includegraphics[width=0.23\textwidth,clip]{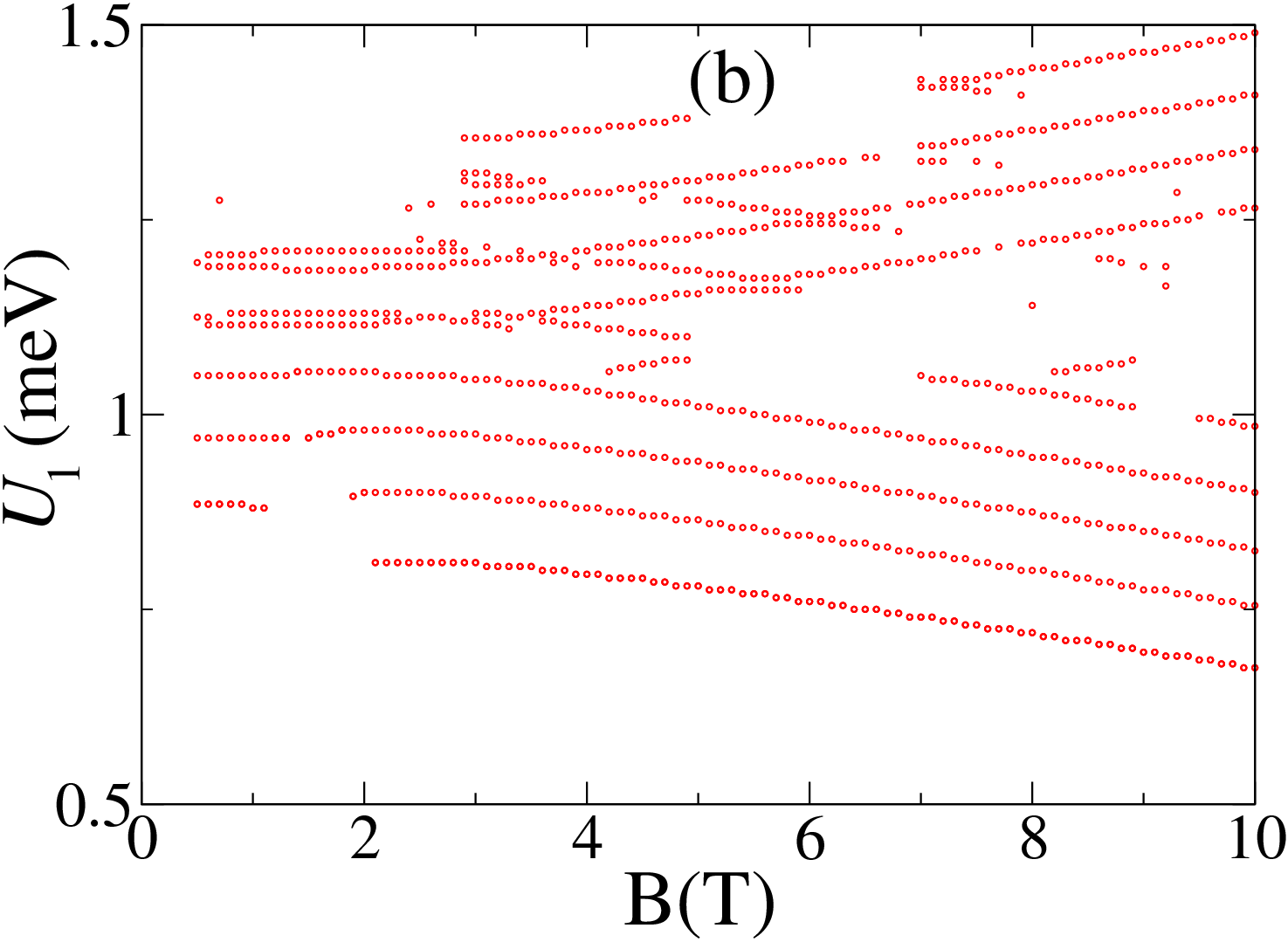}
\caption{\emph{(Color online).
Position of the magnetoconductance peaks as a function of magnetic field for parallel (a) and 
perpendicular (b) orientations of the fields.}}
\label{fig:peaks}
\end{figure}

We have solved the scattering problem using numerical integration of the Schrodinger equation via the 
fourth-order Adams-Moulton method, which was required due to the presence of the first power 
of momentum in the Hamiltonian (\ref{simpHamil}). The conductance due to a single transverse channel, 
according to the Landauer formula, is related to the $2\times 2$ spin-dependent transmission 
amplitude $A$ by
\begin{gather}
G = \frac{e^2}{h}P(\epsilon)\ , \ \ \ P(\epsilon)= \text{tr} A^\dagger A \ .
\end{gather}
The calculated conductance for fields 5T, 10T,  and 15T in the parallel and perpendicular 
orientations is plotted in Fig.\ref{fig:conductance} versus $U_1$, which is the contribution to the effective
1D potential which is constant between the barriers defined in (\ref{U2}).
The difference between the resonant structure observed for the two field orientations is due to the Dresselhaus interaction.
Whereas the interaction does not influence the resonant pattern for the case of parallel field, it does so for the perpendicular
orientation. We plot the field dependence of the resonant peak positions in Fig.\ref{fig:peaks}.
For a parallel applied field, the positions of the peaks are shifted linearly in the magnetic field, 
with slope equal to the magnetic moment $\pm \frac{1}{2} g^*\mu_B$ and opposite spins exhibiting shifts in 
opposite directions. The situation for perpendicular field, however, is markedly different: in the low field 
regime the positions of the peaks depend non-linearly on the applied field and evolve into straight lines 
with slope $\pm \frac{1}{2} g^*\mu_B$ as the field is increased. In the high field regime, the positions of 
the peaks are still off-set relative to those in the parallel orientation, even though spin dynamics is 
becoming adiabatic. This is due to the fact that the Dresselhaus interaction not only induces a non-abelian 
phase when nonadiabatic precession occurs, but also contributes to the abelian phase even in the adiabatic 
regime, so that the peaks in high field remain offset due to the accumulation of a dynamical phase in the 
scattering regions in which spin undergoes significant precession.

\begin{figure}[ht]
\includegraphics[width=0.23\textwidth,clip]{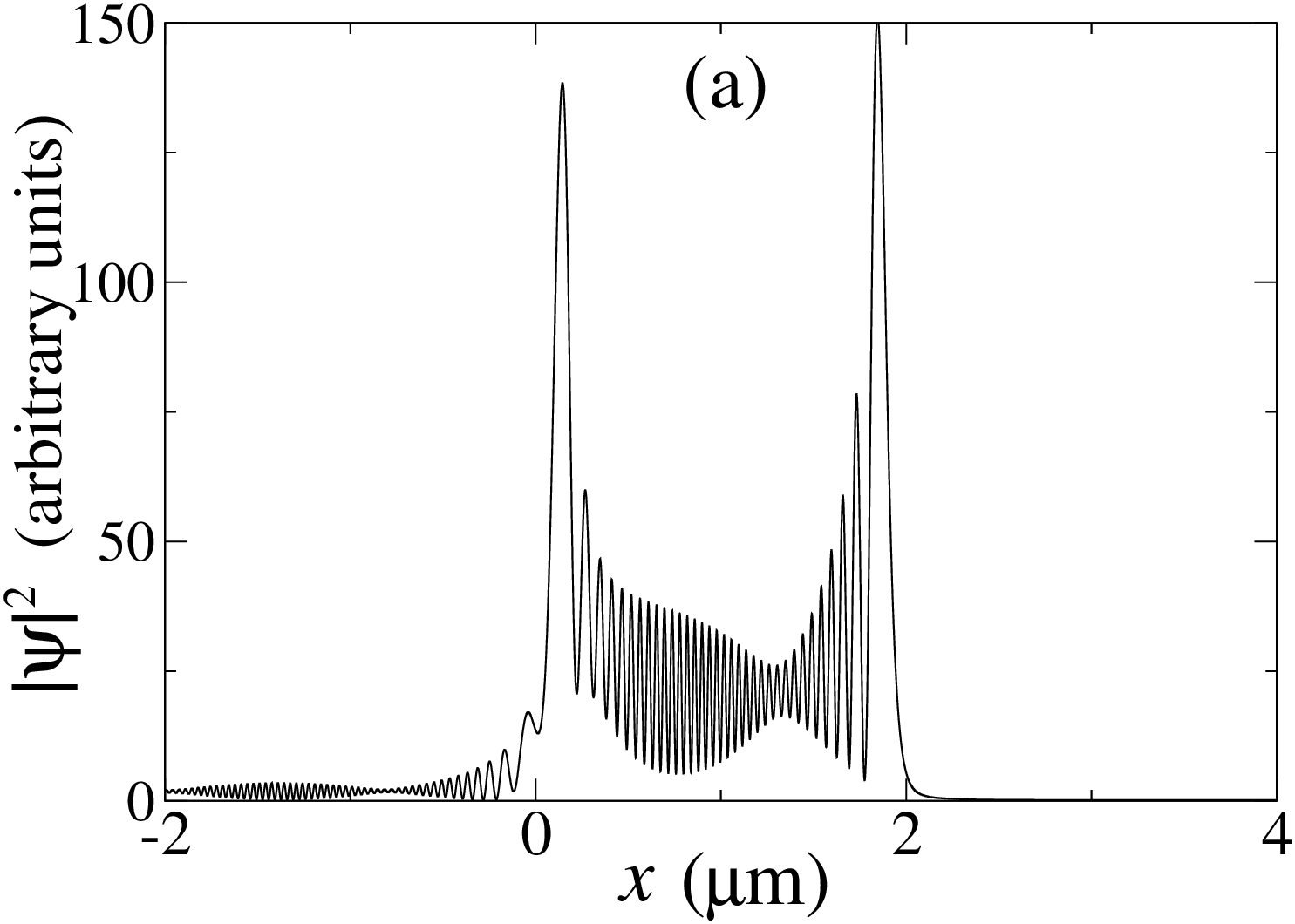}
\includegraphics[width=0.23\textwidth,clip]{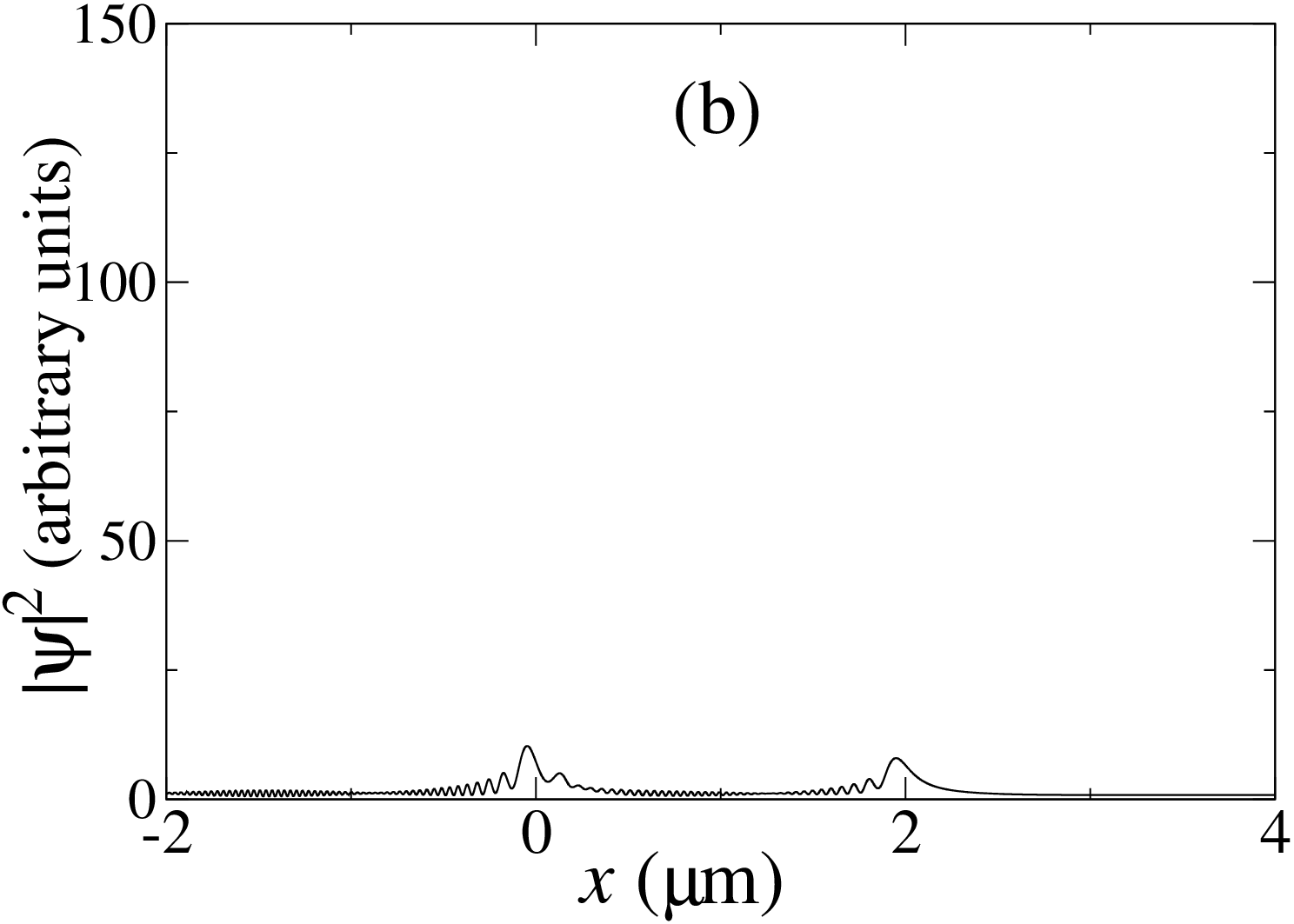}
\includegraphics[width=0.23\textwidth,clip]{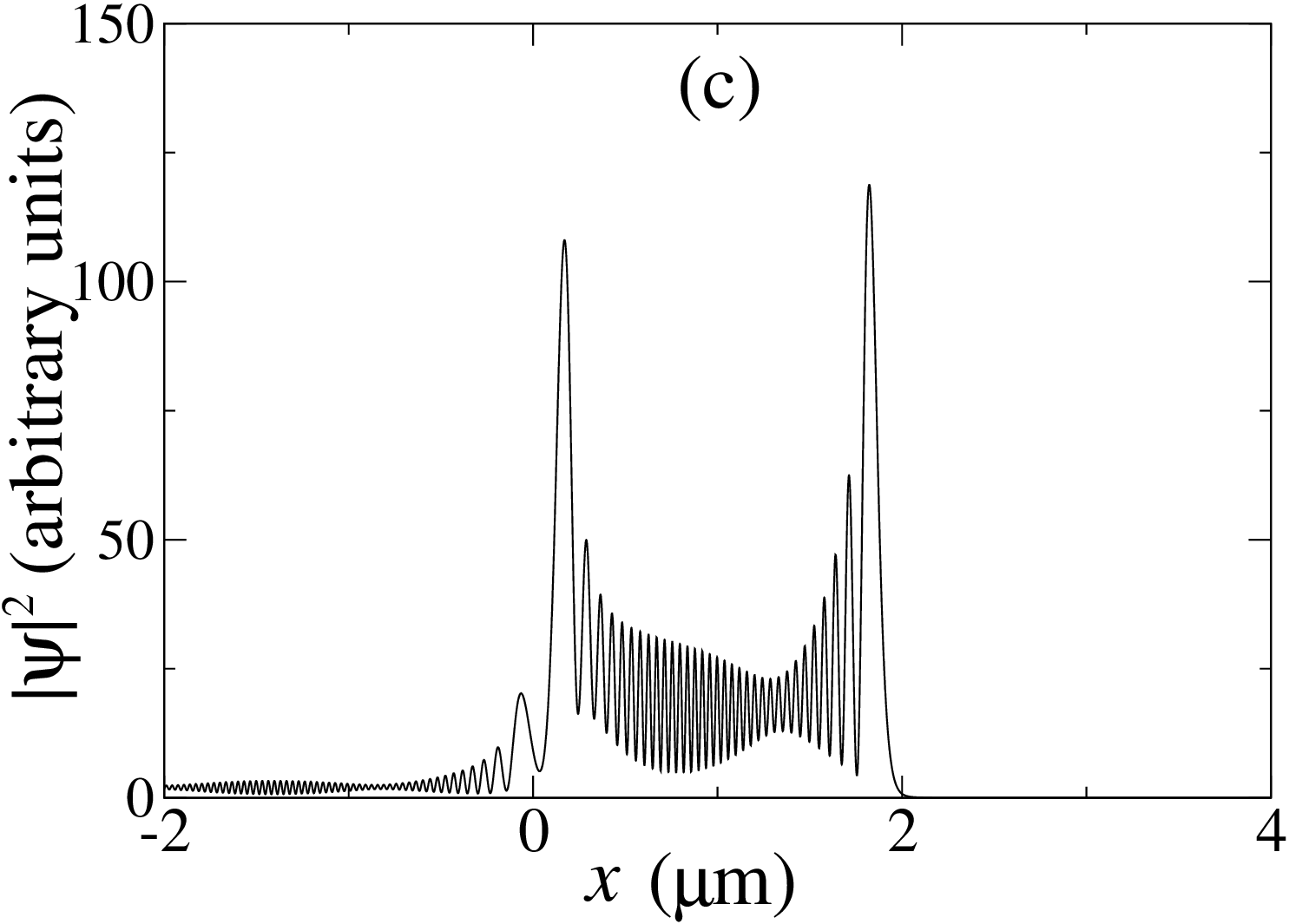}
\caption{\emph{
The probability $|\psi|^2$ plotted on the same scale at three values of $U_1$, corresponding to 
the two {\it negative} resonant cases (A,C) and one off-resonant case (B) indicated in Fig.\ref{fig:conductance}b. 
The wavefunction is strongly enhanced between the barriers at {\it negative} resonances, clearly demonstrating the 
existence of quasi-bound states.}}
\label{fig:boundstate}
\end{figure}

In addition to the non-linear shift in the positive transmission resonant peaks 
we observe a novel feature in the regime where nonadiabaticity of the spin motion is strong,
these are the {\it negative} resonances appearing on the middle plateau. 
The negative resonances marked by letters (A) and (C) are evident in Fig.\ref{fig:conductance}b
at 10T, and are suppressed as the magnetic field is increased into the adiabatic regime. 
Recall that in the spinless situation, sharp peaks appear in the transmission corresponding resonant transparency 
at energies where a quasi-bound state exists between the barriers. It is clear that appearance of negative peaks 
corresponding to resonant enhancement of reflection cannot exist for parallel fields, and is therefore closely 
tied to non-trivial spin dynamics.  
In Fig.\ref{fig:boundstate} we display the electron wave functions corresponding to energies (A), (B), and (C)
marked in the conductance plot Fig.\ref{fig:conductance}b.
We observe that the wavefunction at energies corresponding to {\it negative} resonances (A) and (C)
is strongly enhanced between the barriers signalling the presence of a quasi-bound state at the two resonant energies.
We reiterate that these states appear at 10T and are associated with a suppression in the transmission.
At high magnetic field the states gradually disappear; they are evident only vestigially at 15T, Fig.\ref{fig:conductance}c.

Practically, the non-linear field dependence of the splitting of ususal (``positive'') transmission resonances  
shown in  Figs. ~\ref{fig:conductance} and ~\ref{fig:peaks} might provide a robust way to probe and to measure
the Dresselhaus and the Rashba interactions. On the other hand observation of the negative resonances might be 
a challenge because of their relatively small spectral weight (we observe numerically that the negative 
resonances are more pronounced for heavy holes than for electrons). On the other hand the negative resonances are 
a qualitatively new feature related to the non-abelian and non-adiabatic spin dynamics and we shall explain the
physical mechanism behind these features in the next section.

\section{The physical  mechanism for negative resonances.}
 Let us consider the region very near the top of the barrier, in which $p_x \approx 0$. Since the 
external magnetic field is dominant near the barrier, the scattering of a state which is incident 
on the single barrier is completely described by the spin-dependent $T$-matrix for a single 
parabolic barrier with \emph{constant} magnetic field,
\begin{gather}
\hat{T} = \left(\begin{array}{cc} \hat{t}_{11} & \hat{t}_{12} \\ \hat{t}_{21} & \hat{t}_{22} \end{array}\right) \ .
\end{gather}
Hereafter we indicate the explicit inclusion of spin by the hat above the $T$-matrix; the dimension of $\hat{T}$ is hence
 4$\times$4. The $2\times 2$ matrices $\hat{t}_{ij}$ are diagonal in the basis of states with spin 
directed along the external field $\vec{B}$,
\begin{gather}
\hat{t}_{ij} = \left(\begin{array}{cc}
t_{ij} ( \epsilon +\epsilon_B) & 0 \\
0 & t_{ij} (\epsilon -  \epsilon_B)
\end{array}\right)\ .
\end{gather}
Here we have written $\epsilon_B = |\frac{1}{2} g^*\mu_B B|$ and the spin-independent matrix 
elements $t_{ij}(\epsilon)$ were given in Eq. (\ref{tmatrix}). In the situation where 
$ 2\epsilon_B \gg \omega_x$, the Zeeman splitting may be clearly resolved, a middle plateau exists, 
and for energies lying on this plateau, each barrier acts as a spin filter, preferentially 
reflecting spins which are aligned with the magnetic field and transmitting spins which are anti-aligned.

Away from the potential barriers, the electron momentum is large and hence
it is possible to perform a Born-Oppenheimer separation of 
orbital and spin motion. Writing the wave-function as
\begin{gather}
\psi(x) = e^{i \int{ k dx}} \chi(x) \ ,
\end{gather}
where $\chi$ is a spinor and
\begin{gather}
k = \sqrt{2m(\epsilon - U)} \ ,
\end{gather}
we find that $\chi$ obeys the following Schrodinger equation
\begin{gather}
\label{etau}
i \frac{d\chi}{d\tau} = -(\epsilon_B \sigma_y + \alpha k \sigma_x) \chi \ .
\end{gather}
Here $\tau$ is an \emph{effective time} defined by
\begin{gather}
\frac{k}{m} \frac{d}{dx} = \frac{d}{d\tau}  \ .
\end{gather}
When deriving Eq. (\ref{etau}) from Eq. (\ref{simpHamil}) one has to remember that between the barriers, 
$\alpha \ll \frac{k}{m}$.
The solution $\chi(\tau)$ may be written in terms of the SU(2) evolution operator $U(\tau)$,
\begin{gather}
\chi(\tau) = U(\tau) \chi(0)
\end{gather}
We have used the same letter for the evolution operator as for the potential, however the meaning should be clear.

In the region between the two barriers, the wave-function consists of counterpropagating waves which 
carry precessing spin. The spins which propagate in opposite directions are related by an operation 
corresponding to the reversal of 'effective time', which differs from the usual time-reversal operator 
in that $\sigma_x$ does not change sign since the $x$-component of the effective magnetic field 
$\alpha p_x$ is also reversed. It is therefore necessary to augment the time-reversal operator by a 
rotation of $\pi$ about the $y$-axis. The unitary evolution operator then obeys the relation
\begin{gather}
U(-\tau) = e^{ - \frac{ i\pi \sigma_y }{2} } U(\tau)^\dagger e^{ \frac{ i\pi \sigma_y}{2}} \ .
\end{gather}
Since the region in which the electron sees a relatively constant magnetic field extends over a large 
number of de Broglie wavelengths, while the scattering region consists of a small number of de Broglie 
wavelengths, the $T$-matrix for the complete process is simply given by the product of the individual 
$T$-matrices at each barrier with the matrix describing phase and spin evolution between the barriers,
\begin{gather}
\hat{T}' = 
\hat{T}
\left(\begin{array}{cc}
e^{i \Phi} U(\tau) & 0 \\
0 &e^{-i \Phi} U(-\tau)
\end{array}\right)
\hat{T}
\end{gather}
where $e^{i \Phi}$ is an abelian phase and
\begin{gather}
U(\tau) = \left(\begin{array}{cc}
\alpha & -\beta^* \\
\beta & \alpha^*
\end{array}\right), \ \ \ U(-\tau) = \left(\begin{array}{cc} 
\alpha^* & -\beta \\
\beta^* & \alpha
\end{array}\right)
\end{gather}
The spin-dependent 2$\times$2 transmission amplitude is given by
\begin{gather}
A = \hat{t}_{11}^{\prime -1} = ( \hat{t}_{11} e^{i \Phi} U(\tau) \hat{t}_{11} + 
\hat{ t}_{12} e^{-i \Phi} U(-\tau) \hat{t}_{21} )^{-1}
\end{gather}
When the energy lies on the middle plateau, $\epsilon + \epsilon_B \approx \epsilon_B \gg 0, \epsilon - \epsilon_B \approx -\epsilon_B \ll 0$, we obtain, making use of the explicit forms (\ref{tmatrix})
\begin{widetext}
\begin{eqnarray}
&&\hat{t}'_{11}\approx\nonumber\\ 
&&-e^{i \Phi} 
\left(\begin{array}{cc} 1 & 0 \\ 0 & \sqrt{ 1 + e^{ 2\pi \epsilon_B/\omega_x} } \end{array}\right)
\left(\begin{array}{cc}
\alpha & -\beta^* \\
\beta & \alpha^*
\end{array}\right) 
\left(\begin{array}{cc} 1 & 0 \\ 0 & \sqrt{ 1 + e^{ 2\pi \epsilon_B/\omega_x}} \end{array}\right) + 
 e^{-i \Phi} \left(\begin{array}{cc}  0 & 0 \\ 0 & e^{\pi \epsilon_B/\omega_x} \end{array}\right) 
\left(\begin{array}{cc} \alpha^* & -\beta\\ \beta^* & \alpha \end{array}\right)
\left(\begin{array}{cc} 0 & 0 \\ 0 & e^{ \pi \epsilon_B/\omega_x}
\end{array}\right) \nonumber\\
&&= \left(\begin{array}{cc}
- \alpha e^{i \Phi} & -\beta^* e^{ \pi \epsilon_B/\omega_x} e^{i \Phi} \\
\beta e^{ \pi \epsilon_B/\omega_x} e^{i\Phi} & (\alpha e^{-i \Phi} - \alpha^* e^{i \Phi} ) e^{ 2\pi \epsilon_B/\omega_x} - \alpha^* e^{i \Phi}
\end{array}\right) \ .
\end{eqnarray}
The spin-dependent transmission amplitude is
\begin{gather}
A = \frac{e^{-i \Phi}}{ |\alpha|^2  +  e^{2\pi \epsilon_B/\omega_x}(1 - \alpha^2 e^{- 2i \Phi}) } \left(\begin{array}{cc} ( \alpha e^{-2i \Phi} - \alpha^*  ) e^{ 2\pi \epsilon_B/\omega_x} - \alpha^*  & -\beta^* e^{ \pi \epsilon_B/\omega_x} \\
\beta e^{\pi \epsilon_B/\omega} & -\alpha  \end{array}\right) 
\end{gather}
\end{widetext}

The off-diagonal matrix elements $A_{12}, A_{21}$ correspond to transmission with spin-flip. 
The conductance is given by
\begin{gather}
G = \frac{e^2}{h} \text{tr} A A^\dagger = \frac{e^2}{h} (|A_{11}|^2 + |A_{12}|^2 + |A_{21}|^2 + |A_{22}|^2 )
\end{gather}
We may immediately identify the off-resonant situation when the exponential factor $e^{2\pi B/\omega}$ 
is dominant, so that the transmission amplitude and probabilty
are of the form
\begin{eqnarray}
&&A = e^{-i \Phi} \frac{\alpha e^{-2i \Phi } - \alpha^*}{ (1 - \alpha^2 e^{-2i \Phi} )} \left(\begin{array}{cc}
1 & 0 \\
0 & 0
\end{array}\right)\nonumber\\
&&P(\epsilon)=|\alpha|^2\approx 1 \ .
\end{eqnarray}
which corresponds to the filtering of spin at each barrier.

At a usual ``positive'' resonance
\begin{gather}
\label{fullloop}
1 - \alpha^2 e^{-2i \Phi} = 0 \rightarrow \alpha = e^{i (\Phi + n \pi) } \ .
\end{gather}
This implies that $|\alpha|=1$, $\beta=0$.
Hence
\begin{eqnarray}
&&A = \frac{e^{-i \Phi}}{|\alpha|^2} 
\left(\begin{array}{cc}
-\alpha^* & 0 \\
0 & -\alpha
\end{array}\right)\nonumber\\
&&P(\epsilon)=2 \ .
\end{eqnarray}

At a ``negative'' resonance we need to satisfy the following conditions
\begin{eqnarray}
&& \alpha e^{-2i \Phi} - \alpha^*  =0\nonumber\\
&&1 - \alpha^2 e^{- 2i \Phi} \ne 0 \ .
\end{eqnarray}
This is possible only if $\alpha = |\alpha|e^{i (\Phi + n \pi) }$ and
$|\alpha| < 1$. The phase condition is the same as that for the positive
resonace, Eq. (\ref{fullloop}). However, $|\alpha| < 1$  is possible only in a 
nonadibatic case. In this case we have
\begin{eqnarray}
&&A = \frac{e^{-i \Phi}}{|\beta|^2e^{ 2\pi \epsilon_B/\omega_x}}
\left(\begin{array}{cc}
-\alpha^* & -\beta^*e^{ \pi \epsilon_B/\omega_x} \\
\beta e^{ \pi \epsilon_B/\omega_x} & -\alpha
\end{array}\right)\nonumber\\
&&P(\epsilon)=e^{-2 \pi \epsilon_B/\omega_x} \ .
\end{eqnarray}
We see that in order to obtain a negative resonance it is necessary for spin to precess non-trivially, 
so that a lower spin component develops over the course of the trajectory. When motion is significantly 
nonadiabatic, the lower component of spin may undergo resonant enhancement, leading to suppression 
of transmission. When the parameters are driven deeper into the adiabatic regime, we find that the 
resonant behaviour can reverse to become a more common positive resonance, which explains how the negative
spike marked (A) in Fig.\ref{fig:conductance}b appears as a small positive ``bump" on the middle plateau in panel (c) of
the same figure.

In order to observe a negative resonance, it is necessary that nonadiabatic spin dynamics persist 
into the high-field regime, since the Zeeman splitting must be sufficient large to create a middle plateau 
in which one spin channel is filtered. This requires that the Zeeman splitting be larger than the 
longitudinal oscillator frequency, $g^*\mu_B B > \omega_x$, while remaining significantly smaller than the 
Dresselhaus interaction. While this can be accomplished by making the parabolic barrier wider, this 
would also require the distance between barriers to be increased, which is a sensitive issue since the 
size of the system may be required to exceed the ballistic mean free path. We therefore expect that 
while the non-linear splitting of peaks shown in Figs. ~\ref{fig:conductance} and ~\ref{fig:peaks} 
should be confirmed experimentally with relative ease,  realization of the negative resonances may 
provide a challenge. 

\section{Conclusion}
We suggest a spin-orbit interferometer consisting of two QPCs connected in series.
It is shown that due to the Dresselhaus spin-orbit interaction the spin dynamics in the 
interferometer is nonadiabatic in presence of an external magnetic field.
As a result of this nonadiabaticity the positions of the resonant peaks in the transmission are sensitive to the
direction of the magnetic field and the value of the Dresselhaus interaction.
This effect could be used to directly measure the size of the Dresselhaus interaction in a ballistic channel.
While we performed our calculations for an electron system with the Dresselhaus interaction, it is clear that
the same effect exists generically for holes and for the Rashba spin-orbit interaction.

We thank U. Zuelicke, A. R. Hamilton and A. I.  Milstein for stimulating discussions.


\begin{thebibliography}{99}
\bibitem{com2} This conclusion was also confirmed numerically via explicit calculation of the transmission coefficient.
\end{thebibliography}
\end{document}